\documentclass[11pts,twocolumn]{article}
\usepackage[dvips]{graphics}
\usepackage{epsfig}
\usepackage[centertags]{amsmath}
\usepackage{amsfonts}
\usepackage{amssymb}
\usepackage{amsthm}
\usepackage{newlfont}

\begin{document}
\title{Oscillating horizontal bar problem revisited}
\author{
Anindya Kumar Biswas, Department of Physics;\\
North Eastern Hill University, Shillong - 793022\\
email:anindya@nehu.ac.in}
\date{\today}
\maketitle
\begin{abstract}
A simple text book problem in mechanics\cite{klep}, describes a
massive horizontal bar placed on two oppositely rotating rollers,
kept at a fixed center to center distance. Subsequent motion is to
be found out in presence of kinetic frictions at the point of
contacts of the two rollers\cite{demo}. Introducing bulk rolling
friction effects, through the contact planes and considering
viscoelastic rollers, we find that the inverse of the square of
the oscillation frequency of the bar has a linear relationship
with the center to center distance. The gradient and intercept of
the linear relation together with observations about two
consecutive positions of the bar, determine the rolling friction
coefficient of the viscoelastic materials of the two rollers
fully.
\end{abstract}
\begin{section}{Introduction}
An interesting text book problem in the introductory
mechanics(\cite{klep}) is to determine the motion of a bar. The
bar is uniform, hard, heavy and has weight W. It is placed
horizontally on two hard rollers. Both the cylindrical rollers
have got the same radius R. The rollers are mounted such that axes
of the two are parallel. These are rotated with almost the same
constant angular speed $\omega$. The left roller is rotated
clockwise. The right roller is made to revolve anticlockwise. The
axes of rotation are L distance apart. The rollers are made of the
same material. Surfaces are polished to the same degree. Kinetic
friction coefficient is the same for both the rollers at the
contact lines with the bar. Coefficient of kinetic friction,
$\mu_{k}$, is independent of the speeds of the bar and the
rollers. The bar subsequently executes S.H.M.\cite{demo} with
angular frequency $\sqrt{\frac{2g\mu_{k}}{L}}$. The result is
noteworthy. Reciprocal of the square of the angular frequency is
proportional to L. The proportionality constant, apart from 2g, is
$\mu_{k}$. In principle, one can convert this text book problem
into a working principle of an experiment, where the distance
parameter L can be varied with simultaneous measurement of
oscillation frequency, leading to a determination of the
coefficient of kinetic friction $\mu_{k}$.

When the identical rollers are not hard, life tickles into. The
contact lines become planes. Through these planes each roller
transmit one additional couple in order to make the local section
of the bar to rotate along with it. This local torque is
proportional to the normal force at the junction concerned. The
constant of proportionality is the coefficient of rolling
friction. Unlike kinetic friction this depends on speeds of the
rollers as well as of the bar. The bar now oscillates, but
oscillates with a modified frequency. Reciprocal of the square of
the angular frequency is linear in L, not proportional to L. The
straight line representing the linear relation has got an
intercept. The intercept is proportional to the rolling friction
coefficient. This opens up better avenue. Convert the problem with
soft rollers into the working principle of an experiment, where
the distance L is varied with the measurement of oscillation
frequency, leading to the simultaneous determination of the
coefficients of rolling friction as well as kinetic friction.

In the section II, we work out the motion of hard bar on hard
rollers. In the section III, hard bar on soft rollers are dealt
with. In the following section IV, the ensuing experimental avenue
is explored. At the end section V, we discuss about the window in
which one has to restrict oneself, while designing a rolling
friction experiment sophisticating this bar+roller setup.

\end{section}
\begin{section}{Rolling friction zero}
\begin{figure}
\includegraphics{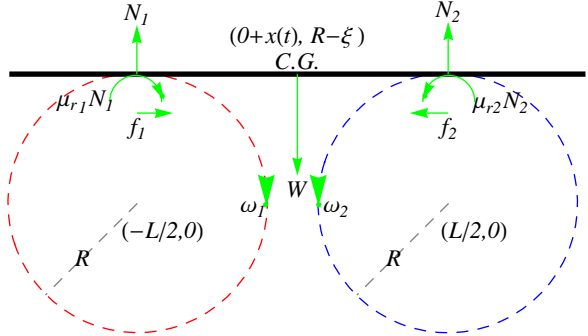}
\caption{A massive hard uniform bar placed on two uniformly
rotating viscoelastic rollers }\label{Figure1}
\end{figure}
As the Fig.\ref{Figure1} shows, five forces are acting on the bar.
The normal force, $N_{1,2}$, is acting through the contact lines
and is perpendicular to the plane of the bar. The weight of the
bar, $W$, is passing through the center of gravity (C.G.)
vertically downwards. The kinetic frictional forces, $f_{1,2}$,
are being exerted by the rollers below, through the lines of
contact, along the directions of motion of the rollers. As the bar
is uniform, the center of mass coincides with the the C.G. $x(t)$
is the position of the center of gravity from the midpoint of the
two rollers at an instant t. The bar is not translating in the
vertical direction. Applying Newton's law along the vertical
direction one gets
\begin{equation}\label{e:for}
N_{1}+N_{2}-W=0
\end{equation}
The bar is not rotating about the center of gravity. Applying
torque equation about an axis parallel to the axes of the rollers,
at the C.G., one is led to
\begin{equation}\label{e:tor}
N_{1}(\frac{L}{2}+x(t))-N_{2}(\frac{L}{2}-x(t))=0.
\end{equation}

The net force on the bar along the horizontal direction is
\begin{equation}\label{e:res}
f_{1}-f_{2}=\mu_{k} N_{1}-\mu_{k} N_{2}.
\end{equation}
provided one assumes that the kinetic friction coefficient,
$\mu_{k}$, is independent of the speed of the bar and the rollers.

Application of the Newton's equation in the horizontal direction
together with eqns.(\ref{e:for}, \ref{e:tor}, \ref{e:res}) leads
one to the differential equation of motion,
\begin{equation}\label{e:diff}
\frac{d^2x(t)}{dt^2} = - \frac{\mu_{k} g}{L}2x(t)
\end{equation}
The eq.\ref{e:diff} implies that the bar executes simple harmonic
motion with all allowable amplitude $A$ and phase $\phi$,
\begin{equation}\label{e:eom}
x(t)= Acos(\omega_{b}t+\phi).
\end{equation}
Angular frequency, $\omega_{b}$, is given by
\begin{equation}\label{e:fre}
\frac{1}{\omega_{b}^{2}}=\frac{1}{2\mu_{k} g}L.
\end{equation}
One notices that the angular frequency, $\omega_{b}$, does not
depend on the mass of the bar. It just depends on the
characteristics of the two roller system. Moreover, as one varies
the separation, $L$, of the two roller system and plot the
resulting $\frac{1}{\omega_{b}^{2}}$ against $L$, as in the
Fig.\ref{Figure3}, overlooking the intercept, the slope gives
$\frac{1}{2\mu_{k} g}$. $\mu_{k}$ is the kinetic friction
coefficient. This might already be an working principle for the
coefficient of kinetic friction measurement.

\begin{figure}
\includegraphics{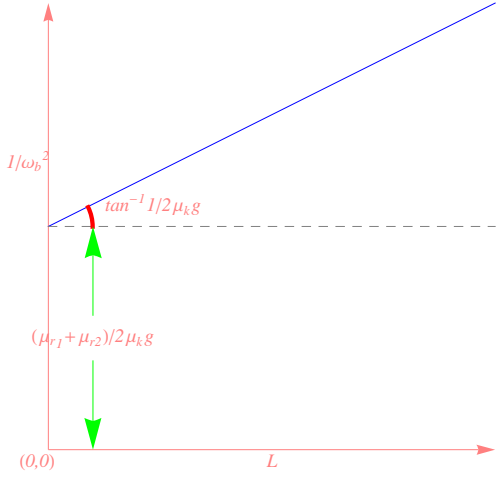}
\caption{predicted plot of $\frac{1}{\omega_{b}^{2}}$ against $L$
}\label{Figure3}
\end{figure}
\end{section}

\begin{section}{Rolling friction non-zero}
Rolling friction is predominantly a bulk effect\cite{brilliant}.
It is due to hysteresis loss during repeated cycles of deformation
and recovery of the body undergoing rolling. Rolling friction
mainly depends upon four parameters. These are mechanical
properties of the body, radius of the rolling object, softness or,
hardness of the body, bar here, in contact and the forward speed
respectively. Mechanical properties involved are elastic constants
and viscosity coefficients.

However, effects of rolling friction here is to introduce
$\mu_{r1,2}N_{1,2}$ as shown in the Fig.\ref{Figure1}. These
couple $\mu_{r1,2}N_{1,2}$ originate, as shown in the
Fig.\ref{Figure2} for the left roller, from the pair of viscous
reaction forces. These viscous reaction couples on the bar
originate as reactions to the couple of dissipative forces acting,
from within, on the rollers, as these are pressed by the bar from
the above.
\begin{figure}
\includegraphics{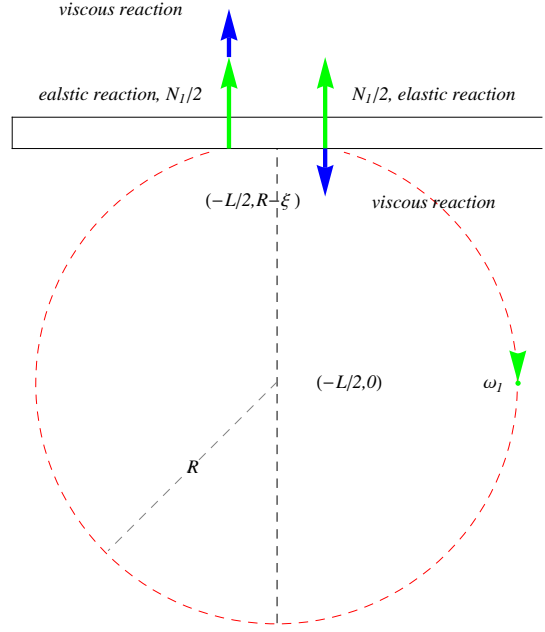}
\caption{$\mu_{r1}N_{1}$ originating from the couple of viscous
reactions on the bar }\label{Figure2}
\end{figure}

As a result, the equation of torque about C.G.(\ref{e:tor}),
changes to,
\begin{equation}\label{e:tor1}
\mu_{r1}N_{1}+N_{1}(\frac{L}{2}+x(t))-(N_{2}(\frac{L}{2}-x(t))+\mu_{r2}N_{2})=0,
\end{equation}
where, $\mu_{r1,2}$ is the rolling frictional torque acting at the
contact planes. Since, the bar is not translating in the vertical
direction, the equation for the net force,(\ref{e:for}), remains
the same. Application of the Newton's equation in the horizontal
direction together with eqns.(\ref{e:for}, \ref{e:tor1},
\ref{e:res}) now leads one to the differential equation of motion,
\begin{equation}
 \frac{d^2x(t)}{dt^2} =- \frac{\mu_{k}
 g}{L+\mu_{r1}+\mu_{r2}}(2x(t)+\mu_{r2}-\mu_{r1})
\end{equation}
Hence, the bar executes simple harmonic motion with all allowable
amplitude $A$ and phase $\phi$,
\begin{equation}\label{e:eom}
x(t)= \frac{\mu_{r1}-\mu_{r2}}{2}+Acos(\omega_{b}t+\phi),
\end{equation}
with modified angular frequency, $\omega_{b}$,
\begin{equation}\label{e:ang1}
\omega_{b}^{2}= \frac{2\mu_{k} g}{L+\mu_{r1}+\mu_{r2}}.
\end{equation}
Ideally, $\mu_{r1}=\mu_{r2}$. Practical limitations may not allow
one to achieve the same angular speed for the two rollers, leaving
$\mu_{r1}=\mu_{r2}$ unattainable.
\end{section}
\begin{section}{Experimental Avenue}
We cast the eq.\ref{e:ang1}, in the more revealing form,
\begin{equation}
\frac{1}{\omega_{b}^{2}}=\frac{1}{2\mu_{k}
g}L+\frac{\mu_{r1}+\mu_{r2}}{2\mu_{k} g}.
\end{equation}
As shown in the Fig.\ref{Figure3}, the intercept is
\begin{equation}\label{e:intercept}
intercept = \frac{\mu_{r1}+\mu_{r2}}{2\mu_{k} g}, slope =
\frac{1}{2\mu_{k} g}
\end{equation}
Moreover, one notices that the equation of motion for the bar
leads to
\begin{equation}\label{e:difference}
x(t)+x(t+\frac{\pi}{\omega_{b}})=\mu_{r1}-\mu_{r2}
\end{equation}
Combination of the two equations (\ref{e:intercept},
\ref{e:difference}) gives $\mu_{r1}$, $\mu_{r2}$ separately.

Again,
\begin{equation}\label{linear}
\mu_{r}=k_{rol}R \omega,
\end{equation}
$R$ and $\omega$ are the radius and angular speed of the roller
concerned. This leads us to $k_{rol}$, velocity independent part
of the velocity dependent\cite{{brilliant},{brilliant1}} rolling
friction coefficient for the roller material. This works as long
as the speed of sound in the roller$\gg R\omega\gg A \omega_{b}$
with $\frac{\xi}{R\omega}\gg$ dissipative relaxation time of the
rollers.

\end{section}
\begin{section}{Discussion}
Differences in dip, $\xi$, as in Fig.\ref{Figure1} of right and
left rollers, due to the difference, $N_{2}-N_{1}$, leads to
changes in the torque and force equations of the order of
$\frac{\xi}{R}$. Radius of the rollers being much larger, this
contribution is neglected. Ideally the experiment should be done
in the small and smaller angular velocity range of the roller and
the bar\cite{pos} so that constancy of the kinetic friction is
maintained. The linear relation for the rolling friction
coefficient(\ref{linear}) is to be replaced by an appropriate
non-linear relation\cite{{xu},{xu1}} when the roller speeds are
higher. Here, we have ignored rolling friction effect due to
surface. See the discussion in second page of the
reference\cite{brilliant}. One can remove little air drag on the
motion of the massive bar by putting the setup in a vacuum
chamber. Finally its nice to notice that such a simple text book
problem leads to an interesting avenue into the experiment of
rolling friction for viscoelastic material. Practical realisation
of this avenue remains something to be done next.
\end{section}

\footnote{Formerly in BITS-Pilani Goa Campus, Vasco, Goa, Pin-403726.\\
email:anindya@bits-goa.ac.in}
\end{document}